\documentclass{PoS}
\usepackage[]{epsfig}

\rightline{\parbox{2.5cm}{\large\rm DESY 05-179 \\HU-EP-05/55 }
\vspace{1cm}}

\PoS{PoS(LAT2005)171}

\title{Probing the chiral phase transition of $N_f=2$ clover fermions with valence overlap fermions}

\ShortTitle{Probing the chiral phase transition with overlap fermions}

\author{\speaker{Volker Weinberg}\\
  John von Neumann-Institut f\"ur Computing NIC, 15738 Zeuthen, Germany\\
  Institut f\"ur theoretische Physik, Freie Universit\"at Berlin, 14196 Berlin, Germany\\
  E-mail: \email{volker.weinberg@desy.de}}

\author{Ernst-Michael Ilgenfritz\\
  Institut f\"ur Physik, Humboldt Universit\"at zu Berlin, 12489 Berlin, Germany\\
  E-mail: \email{ilgenfri@physik.hu-berlin.de}}

\author{Karl Koller\\
  Sektion Physik, Universit\"at M\"unchen, 80333 M\"unchen, Germany\\
  E-mail: \email{karl.koller@lrz.uni-muenchen.de}}

\author{Yoshiaki Koma\\
  Deutsches Elektronen-Synchrotron DESY, 22603 Hamburg, Germany\\
  E-mail: \email{yoshiaki.koma@desy.de}}

\author{Gerrit Schierholz\\
 Deutsches Elektronen-Synchrotron DESY, 22603 Hamburg, Germany\\
 John von Neumann-Institut f\"ur Computing NIC, 15738 Zeuthen, Germany \\
 E-mail: \email{gerrit.schierholz@desy.de}}

\author{Thomas Streuer\\
  John von Neumann-Institut f\"ur Computing NIC, 15738 Zeuthen, Germany\\
  E-mail: \email{thomas.streuer@desy.de}}

\author{\large \sc For the DIK-collaboration}

\abstract{
Overlap fermions are a powerful tool for investigating the chiral and
topological structure of the vacuum
and the thermal states of QCD. 
We study various chiral and topological aspects of the finite temperature phase
transition of $N_f=2$ flavours of  ${\cal O}(a)$ improved
Wilson fermions, using valence overlap fermions as a probe. 
Particular emphasis is placed upon the analysis of the spectral density and the localisation properties of the eigenmodes  as well as on the local structure of topological charge fluctuations in the vicinity of the chiral phase transition.
The calculations are done on $16^3\times8$ lattices generated by the DIK collaboration.

}

\FullConference{XXIIIrd International Symposium on Lattice Field Theory\\
                 25-30 July 2005\\
                 Trinity College, Dublin, Ireland}

\newcommand{\Tr}{\mbox{Tr}}
\newcommand{\Or}{\mbox{${\cal O}$}}

\begin{document}

\section{Introduction}

Since overlap fermions have an exact chiral symmetry on the lattice, 
they are an appropriate tool to investigate the chiral phase 
transition.\footnote{Our results on the vacuum structure for $T=0$ 
quenched configurations are presented in the talk by Y. Koma \cite{Koma:2005} 
at this conference.}
Simulations involving the effects of dynamical overlap sea quarks are just 
becoming possible now, but as they are still numerically extremely demanding 
on realistic lattices, we use a hybrid approach by implementing the overlap 
action for the valence quarks on dynamical configurations 
generated with $N_f=2$ flavours of $\Or(a)$ improved Wilson sea quarks.

As the lattice spacing, $a$, for dynamical fermions is connected to both 
the coupling constant $\beta$ and $\kappa_{sea}$, 
increasing the temperature 
$T=1/(N_4 a)$ across the transition 
is affected by an increase in  $\kappa_{sea}$ at fixed values of 
the coupling, $\beta$, and time-extent, $N_4$, of the lattice.
In order to probe the topological properties of dynamical gauge fields in the 
vicinity of the phase transition, we concentrate 
on the values $\kappa_{sea}$ = 0.1343, 0.1348 and 0.1360 corresponding 
to $T/T_c$ = 0.98, 1.06 and 1.27, respectively, at 
fixed $\beta=5.2$ on $16^3\times 8$ lattices generated by the 
DIK-collaboration \cite{Bornyakov:2004ii}. The lattice spacing
is approximately $a=0.12$ fm.
We use $\Or(200)$ configurations for each value of $\kappa_{sea}$.
The transition point $\kappa_c$ can be identified as the point where the 
Polyakov loop susceptibility \mbox{$\chi=N_s^3(\langle L^2\rangle-\langle L\rangle^2)$}, with the Polyakov loop 
$L(\vec{s})=\frac{1}{3}\Tr \prod_{s_4=1}^{N_4} U_4(\vec{s},s_4)$ , assumes its maximum. 
Recent work of the DIK-collaboration on $16^3 \times 8$ lattices at   
$\beta=5.2$ gave $T_c=212(2)$ MeV, corresponding to $\kappa_c=0.13444(6)$ \cite{Nakamura:2005}.

\section{Density and locality of the eigenmodes of the overlap operator}

\begin{figure}[h]
\begin{center}
\epsfig{file=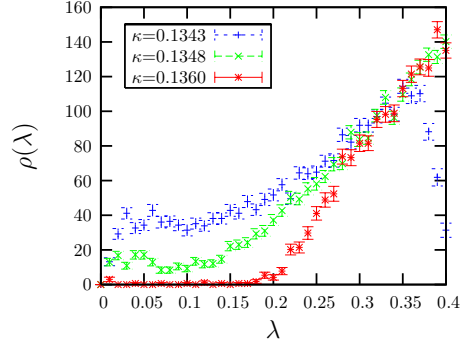,width=6cm}
\end{center}
\vspace{-0.5cm}
\caption[]{The spectral density of the overlap operator in the vicinity 
of the phase transition (50 lowest eigenvalues included).}
\label{fig-density}
\end{figure}

The spontaneous breaking of chiral symmetry by the dynamical creation of a 
nonvanishing chiral condensate, 
$\langle\bar\Psi\Psi\rangle$, 
is related to the spectral density 
$\rho(\lambda)$ of the Dirac operator near zero by the 
Banks-Casher relation 
$\langle\bar\Psi\Psi\rangle=- \frac{\pi}{V} \rho(0)$. 
Using the Arnoldi-algorithm, we compute the 50 lowest eigenvalues on every 
configuration.\footnote{For our implementation of the overlap operator 
see Ref.~\cite{Galletly:2003vf}. Also here we use $\rho=1.4$  and 
stereographically project the eigenvalues.} 
To smooth the configurations we perform 1 step of APE-smearing, using the smearing coefficient $\alpha=0.45$.
In Fig.~\ref{fig-density} we plot the spectral density. One can clearly see a nonvanishing 
chiral condensate in the chirally broken phase  at $\kappa_{sea}=0.1343$  
below the phase transition and a large gap in the spectrum at 
$\kappa_{sea}=0.1360$  in the symmetry restored phase.
However, at $\kappa_{sea}=0.1348$, a value which was determined by the 
Polyakov-loop method to be above the transition, we find a nonvanishing 
tail of small eigenvalues extending down to zero.

\begin{figure}[h]
\begin{center}
\begin{tabular}{cc}
\epsfig{file=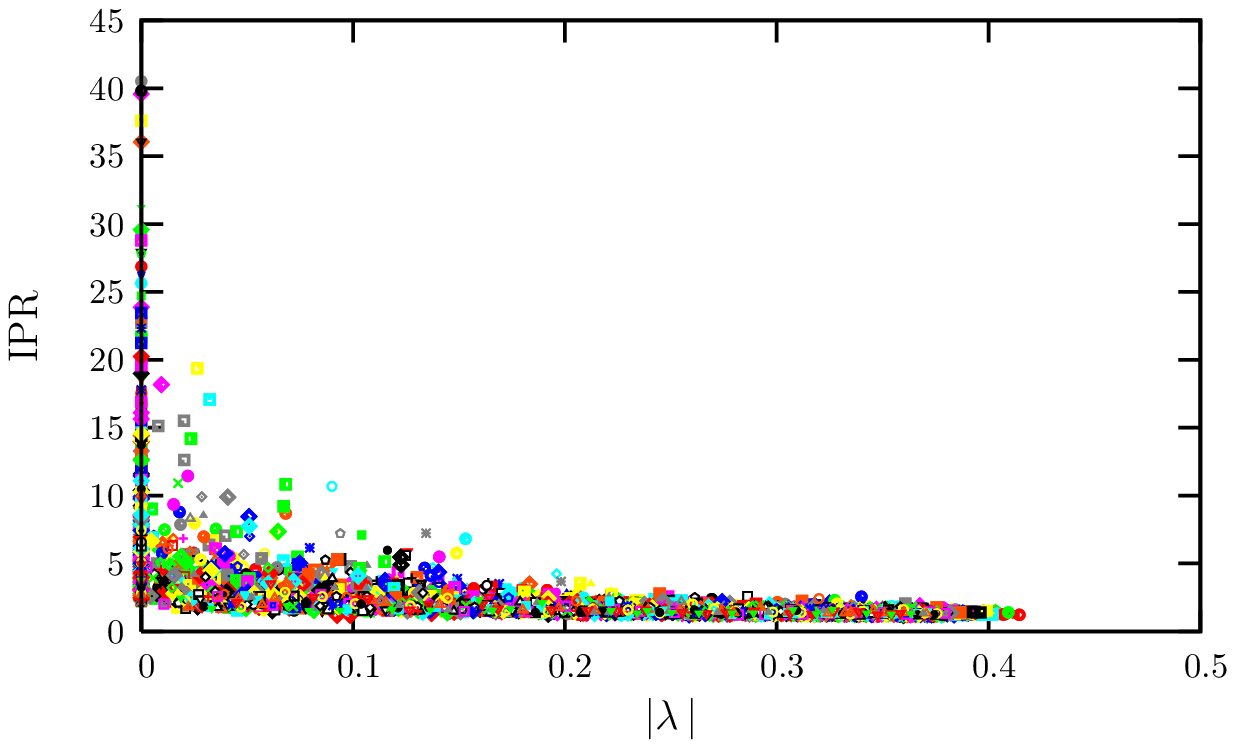,width=6cm}&\epsfig{file=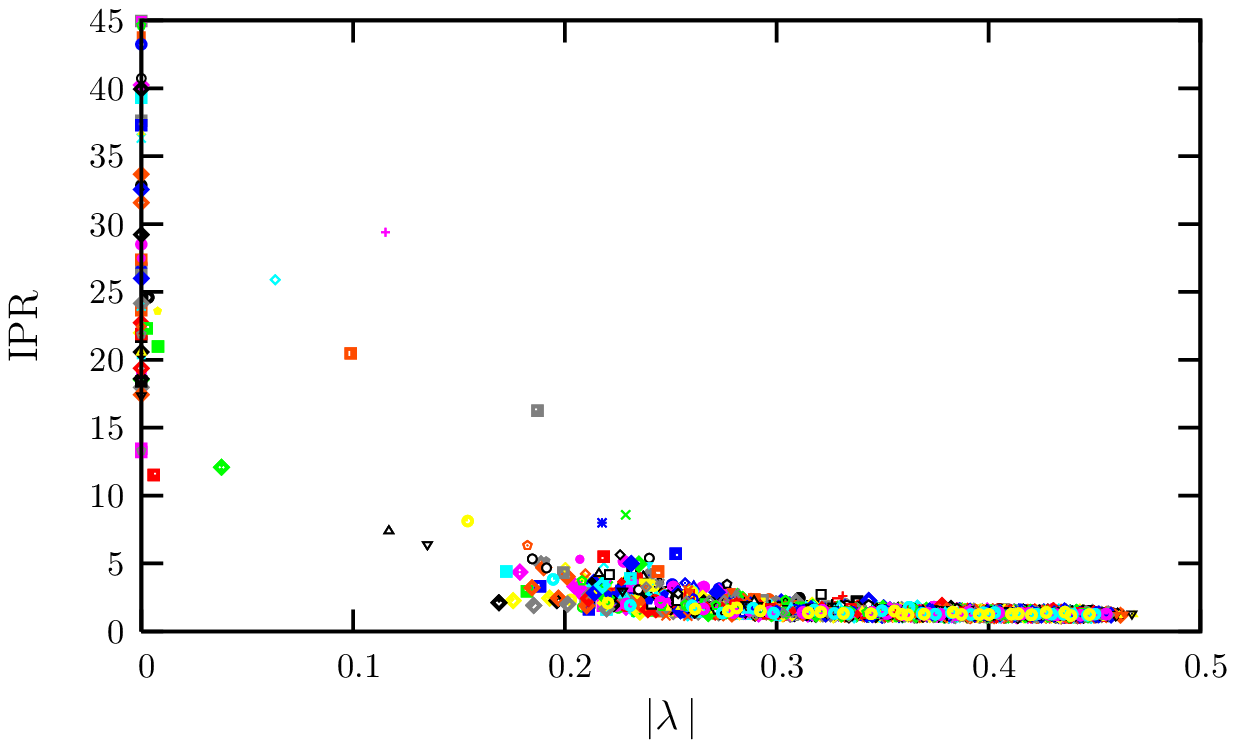,width=6cm}\\
$\kappa_{sea}=0.1343 \;(T<T_c)$ & $\kappa_{sea}=0.1360 \;(T>T_c)$
\end{tabular}
\end{center}
\caption{The IPR of overlap eigenmodes  
(50 eigenmodes from all configurations)
below and above the transition.}
\label{fig-ipr}
\end{figure}

A useful measure  to quantify the localisation of eigenmodes \cite{Aubin:2004mp} is the
inverse participation ratio $I=V\sum_x\rho(x)^2$, 
with the scalar density 
$\rho(x)=\psi_\lambda^\dagger(x)\psi_\lambda(x)$ for
normalised eigenfunctions $\sum_x\rho(x)=1$. 
While $I=V$ if the density has support only on one lattice point, 
$I$ decreases as the density becomes more delocalised, 
reaching $I=1$ when the density is maximally spread on all lattice sites. 
Fig.~\ref{fig-ipr} shows the IPR's for all configurations as a function of 
the modulus of the eigenvalue below and above the phase transition. 
One can see that in both cases the zero modes and the first non-zero eigenmodes are highly localised, 
while the mean IPR for the highest computed eigenvalues is approximately 1.3. 

\section{The local structure of topological charge fluctuations}

The topological charge density for any $\gamma_5$-Hermitean Dirac operator 
satisfying the Ginsparg-Wilson relation is defined as \cite{Hasenfratz:1998ri}:
\begin{equation}
q(x)=\frac{1}{2}\Tr\;\gamma_5\,D(x,x), \;\;Q=\sum_x q(x).
\label{eq-q}
\end{equation}

To compute the topological charge density, we use two different approaches \cite{Horvath:2002yn, Horvath:2003yj}.
In the first approach, we  directly calculate the trace of the overlap operator 
according to equation Eq.~(\ref{eq-q}).
This is a computationally very demanding task and is
therefore performed on only  5 configurations for each $\kappa_{sea}$. 
The ``full'' density computed in this way includes charge fluctuations at 
all scales. The second technique involves the  computation of the topological charge density based 
on ${\cal O}(50)$ low lying modes of the overlap Dirac operator. Using 
the spectral representation of the Dirac operator, the eigenmode expansion
of the topological charge density reads

\begin{equation}
q_\lambda(x)=-\sum_\lambda(1-\frac{\lambda}{2})\; c^\lambda(x),\;\; 
c^{\lambda}(x) =\psi^\dagger_\lambda(x) \gamma_5 \psi_\lambda(x)\;.
\end{equation}

Truncating the expansion acts as an ultraviolet filter by removing the short-distance fluctuations from $q(x)$. Note that the topological charge $Q=\sum_x q_\lambda(x)$ is not affected by the level of truncation.

Reflection positivity, i.e. positivity of the metric in Hilbert space, 
demands that the topological charge correlator is negative \cite{Seiler:2001je} 
\begin{equation}
\label{eq-seiler}
C_q(r)= \frac{1}{V} \sum_x  \langle q(x)q(y)\rangle \le 0, \;\; r>0 \:,
\end{equation}

\noindent where $r=|x-y\,|$ is the Euclidean distance. Since overlap fermions are not ultralocal, the fermion action is not
strictly reflection positive. Therefore,
analysing dynamical or quenched lattices with
valence overlap fermions,
one may expect a positive core and a negative tail of $C_q(r)$~\cite{Koma:2005,Horvath:2005cv}. This will be washed out when a finite UV filter is applied.

\begin{figure}
\begin{center}
\epsfig{file=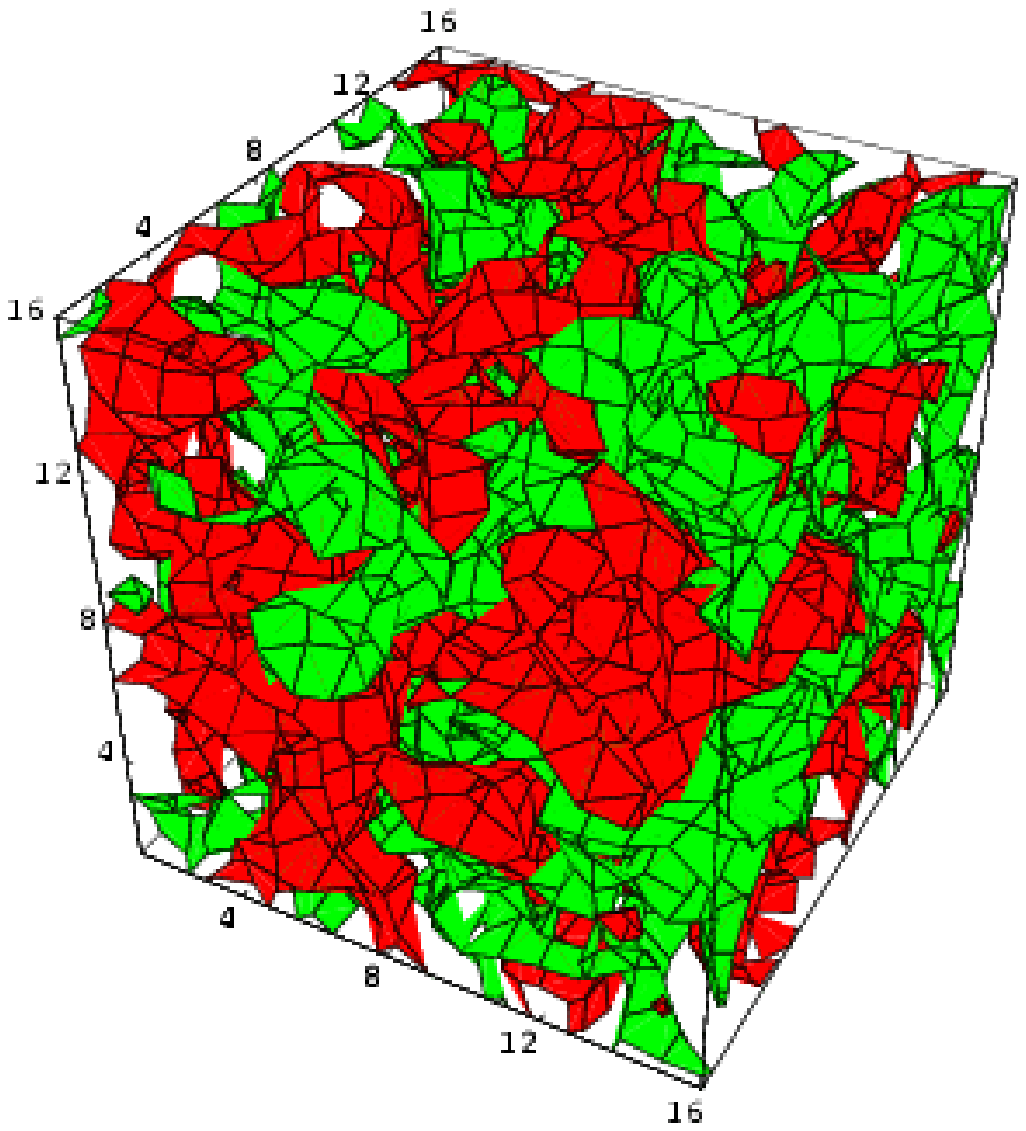,width=3.1cm}\hspace{0.3cm}   
\epsfig{file=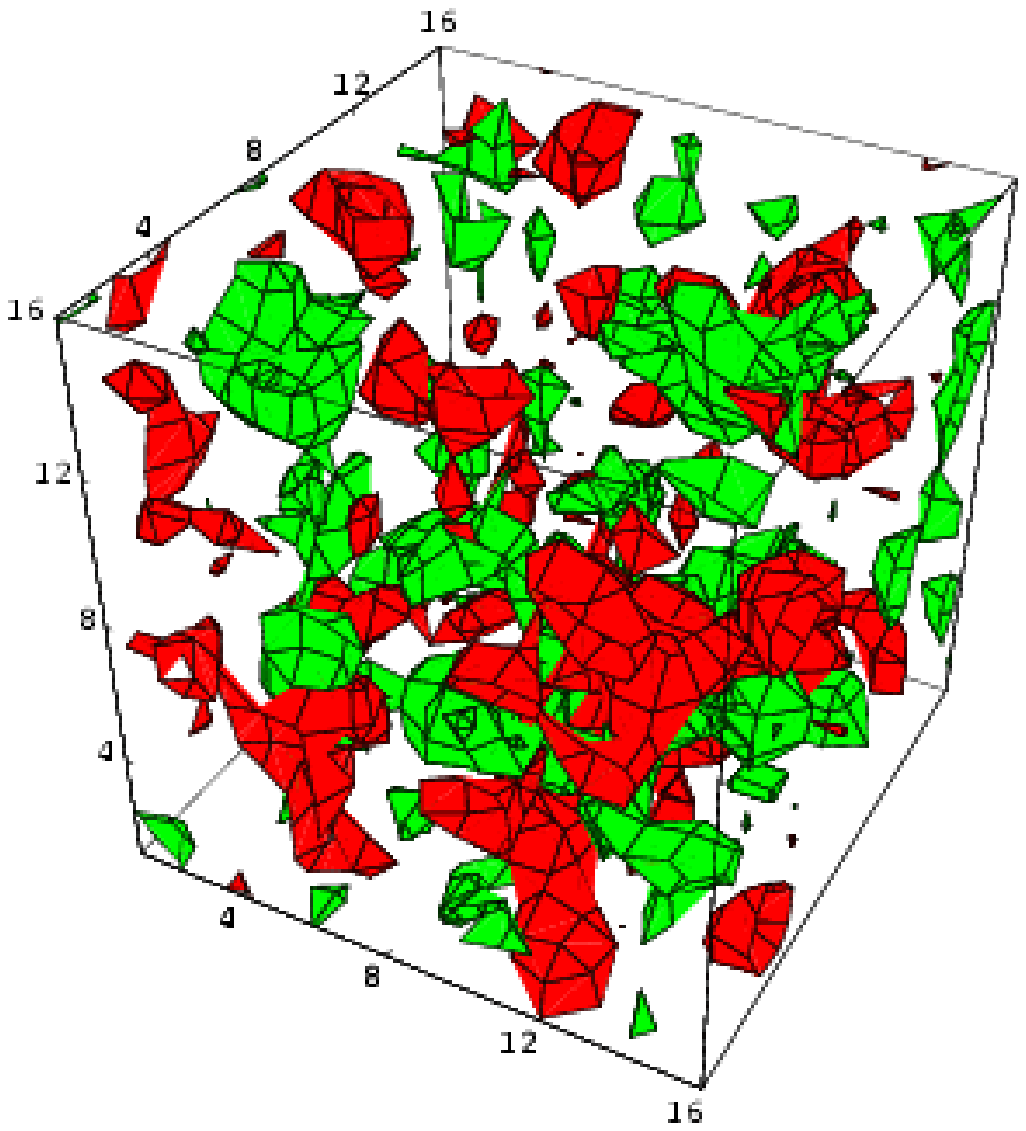,width=3.1cm}\hspace{0.3cm}  
\epsfig{file=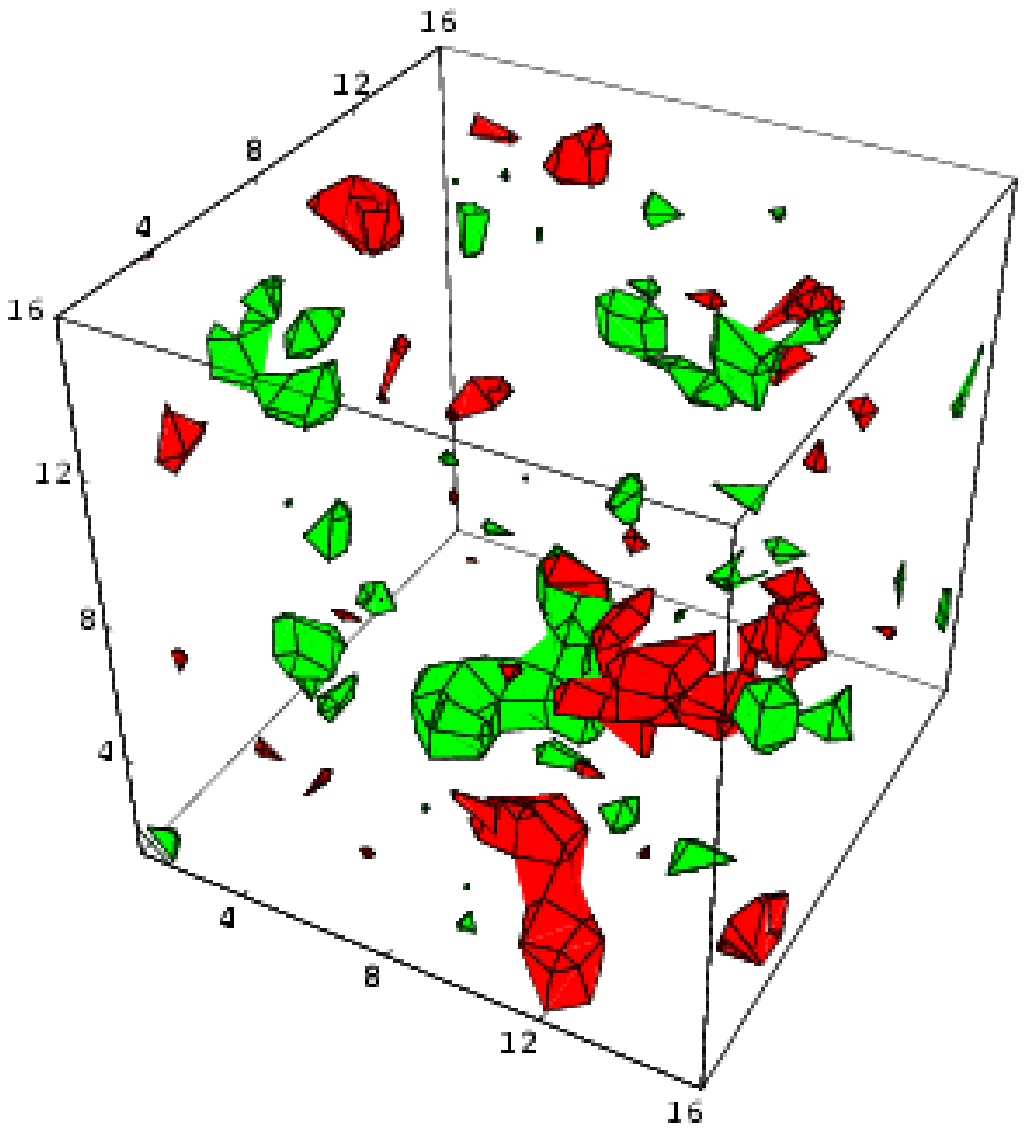,width=3.1cm}\hspace{0.3cm}   
\epsfig{file=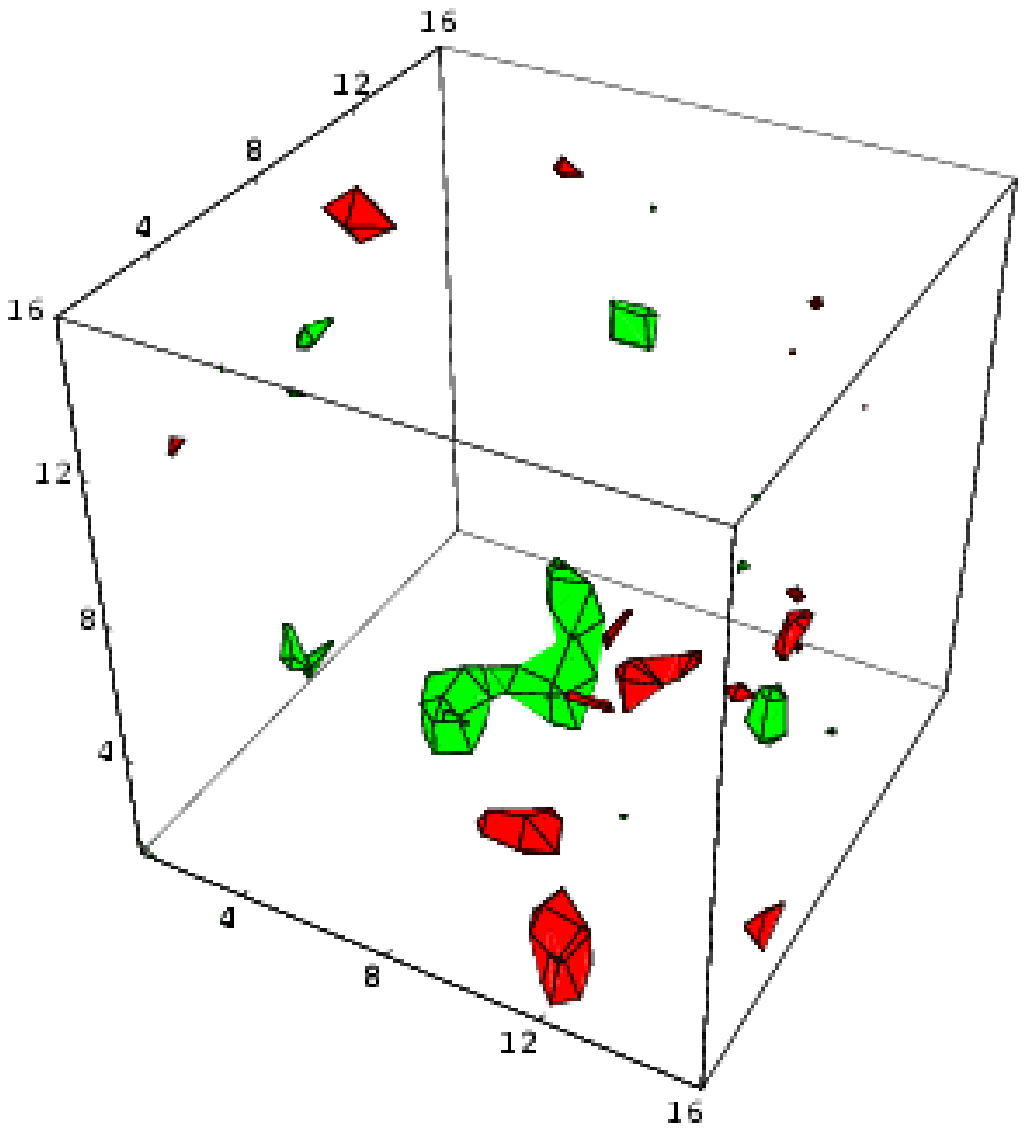,width=3.1cm}\hspace{0.3cm} \\ 
\epsfig{file=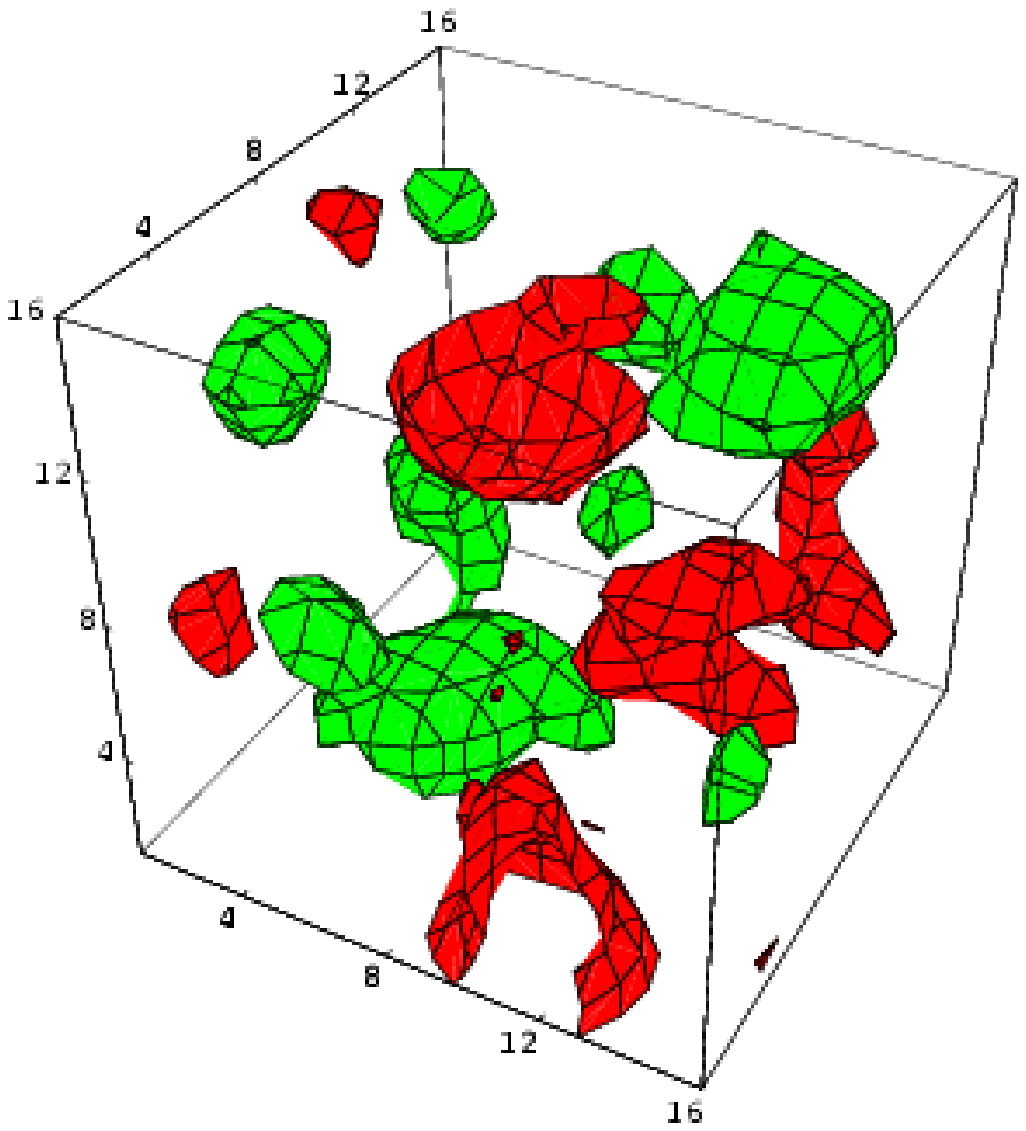,width=3.1cm}\hspace{0.3cm} 
\epsfig{file=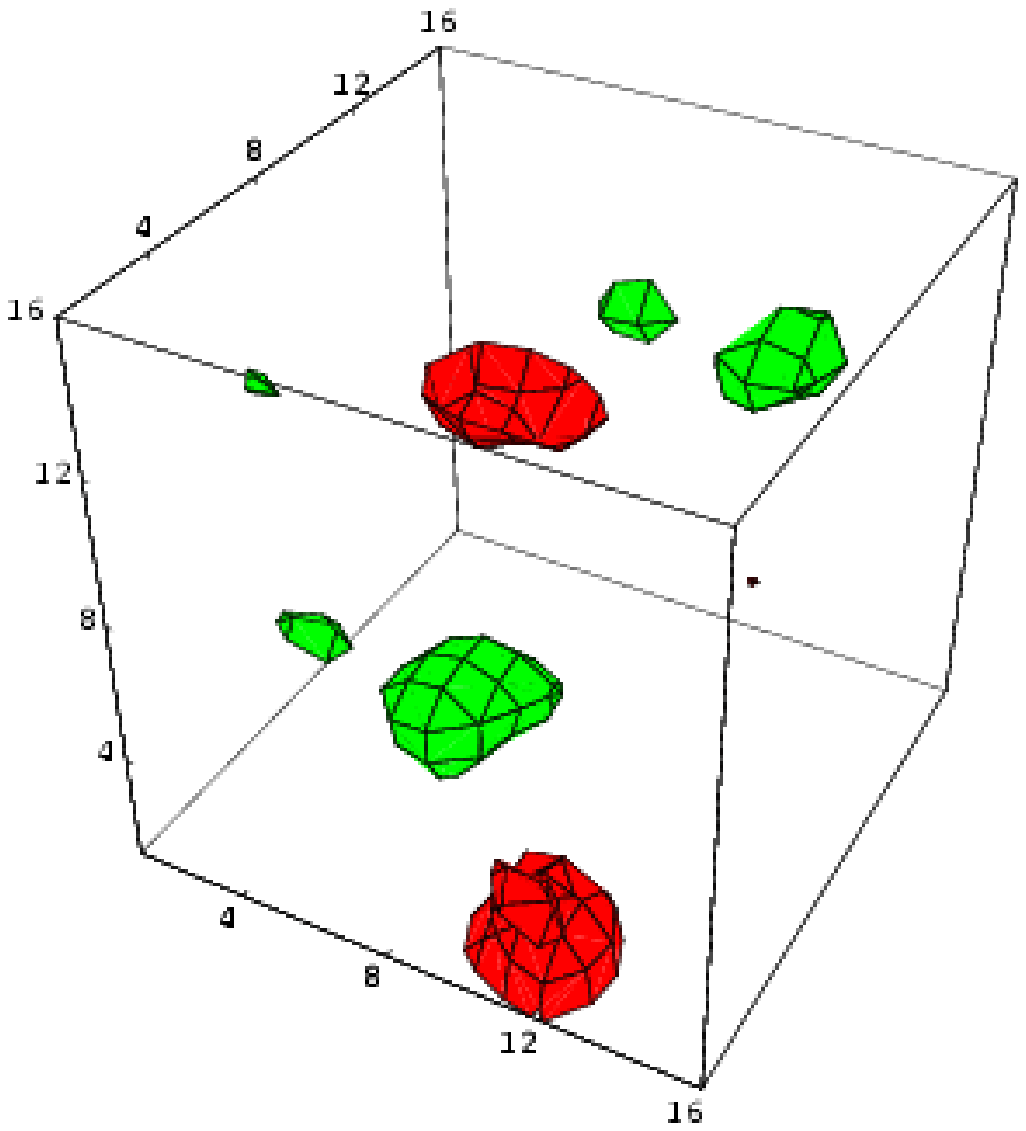,width=3.1cm}\hspace{0.3cm}  
\epsfig{file=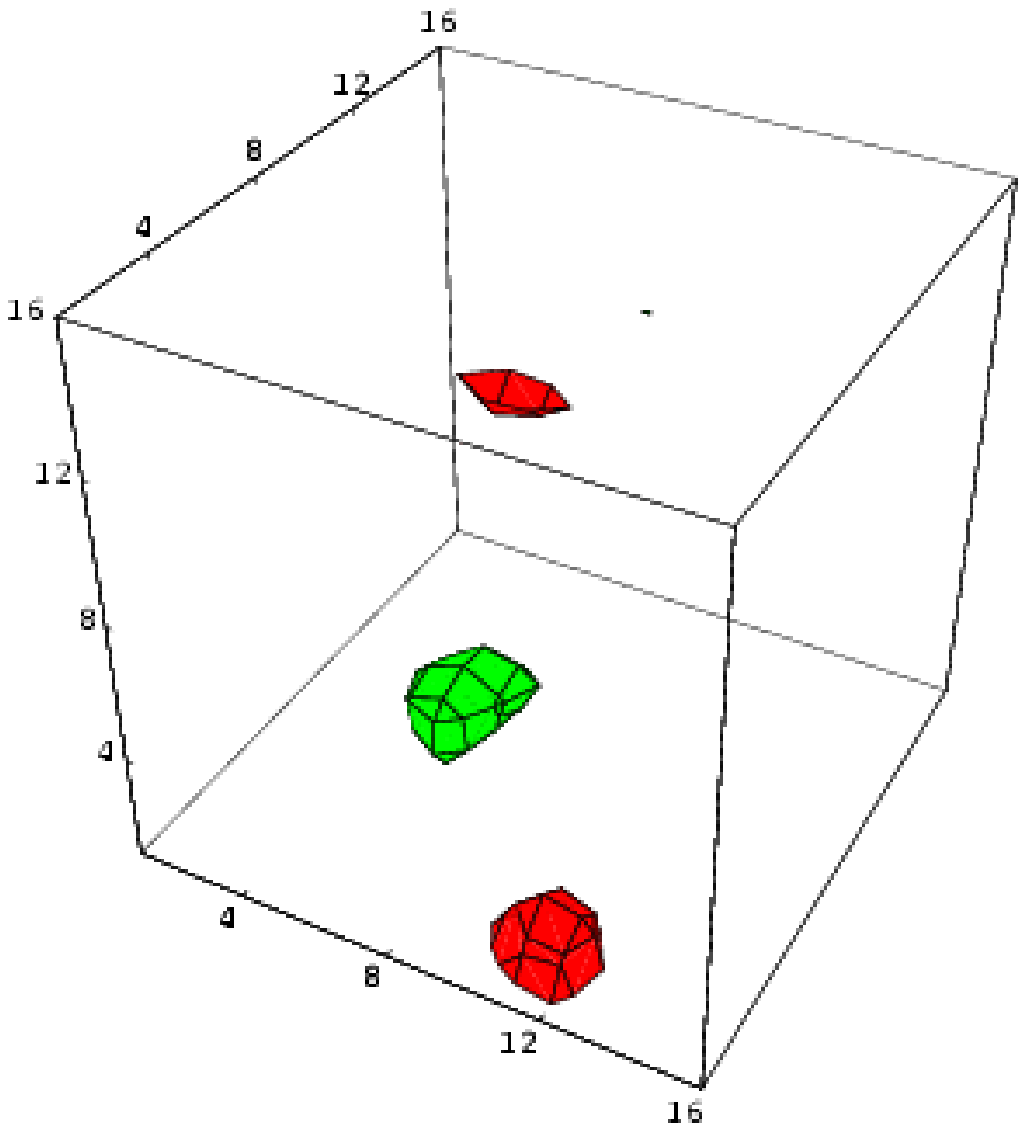,width=3.1cm}\hspace{0.3cm}  
\epsfig{file=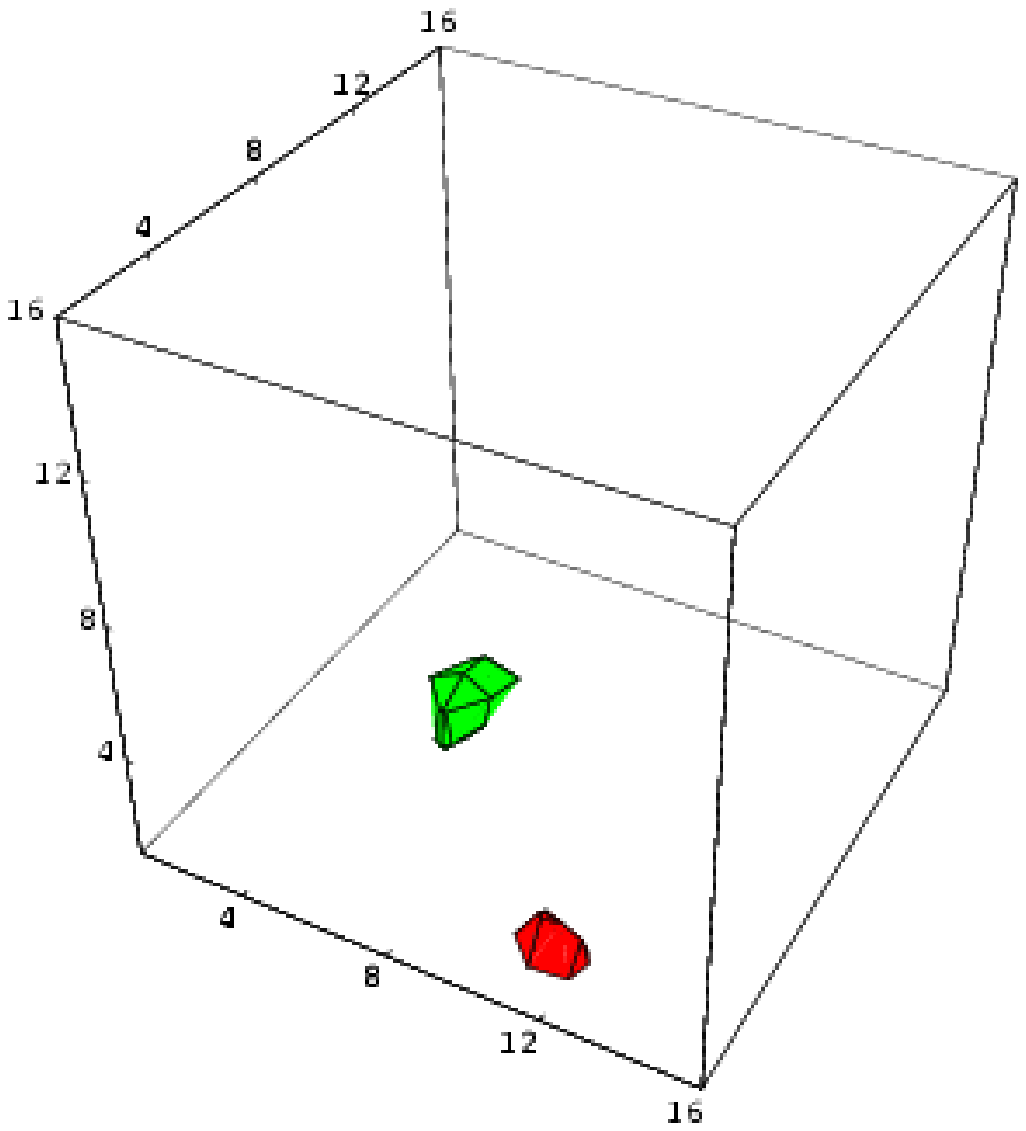,width=3.1cm}

\caption{Isosurfaces of topological charge density with 
$|q(x)|/q_{max} = 0.1$, 0.2, 0.3, 0.4, with red (green) surfaces indicating  
positive (negative) charges 
in one timeslice of a configuration with
$\kappa_{sea}=0.1343$.
The upper (lower) pictures are based on the full (eigenmode truncated) density.
}
\label{fig-contour}
\end{center}
\end{figure}

Let us first consider the local structure of the topological charge
fluctuations, studying sign coherent clusters formed by connected 
neighbours $x_i$ with $|q(x_i)| > q_{cut}  = 0.1 \dots 0.4\: q_{max} $, where $q_{cut}$ is a 
cutoff value to be varied at will.
To give a picture of the spatial distribution of the topological charge density, isosurface plots with $q(x)=\pm q_{cut}$ are shown in Fig.~\ref{fig-contour}.\footnote{Movies are available at  
[\href{http://www.cip.physik.uni-muenchen.de/~weinberg/topdens/}{{\tt http://www.cip.physik.uni-muenchen.de/\~{}weinberg/topdens/}}].}

\begin{figure}[h]
\begin{center}
\hspace*{-0.5cm}\begin{tabular}{cc}
\epsfig{file=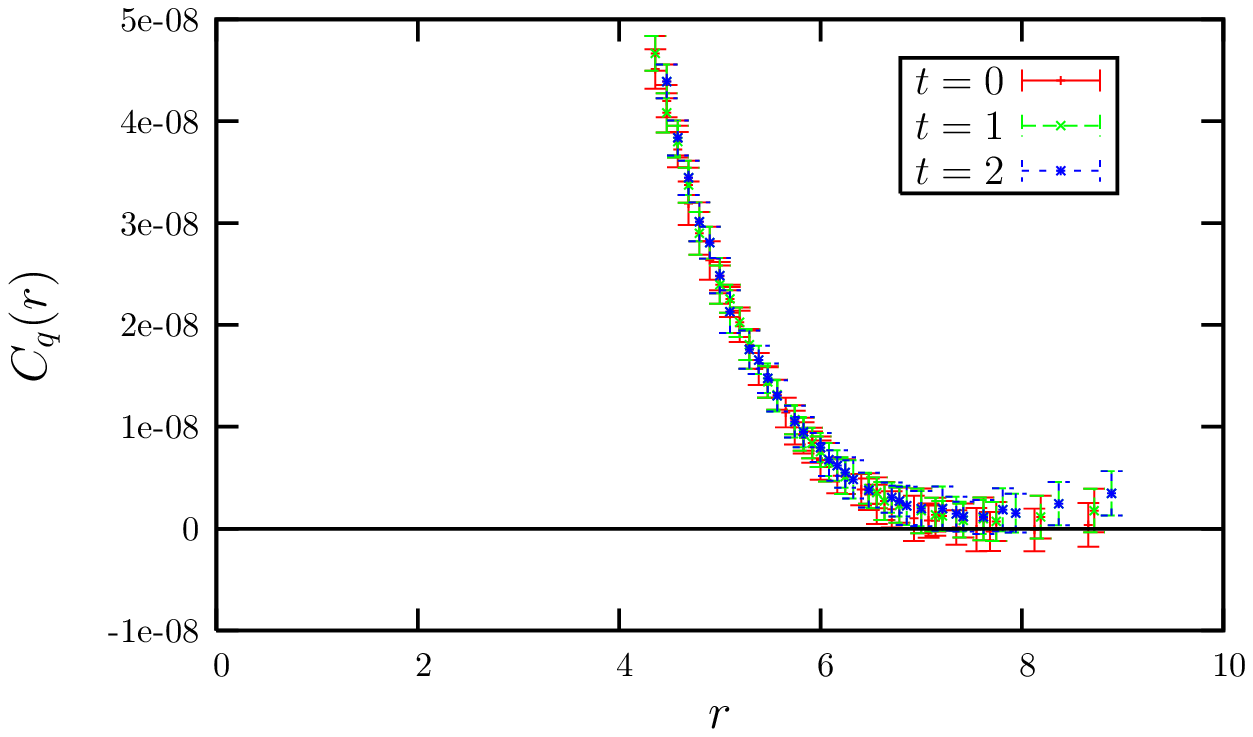,width=7.1cm}&\epsfig{file=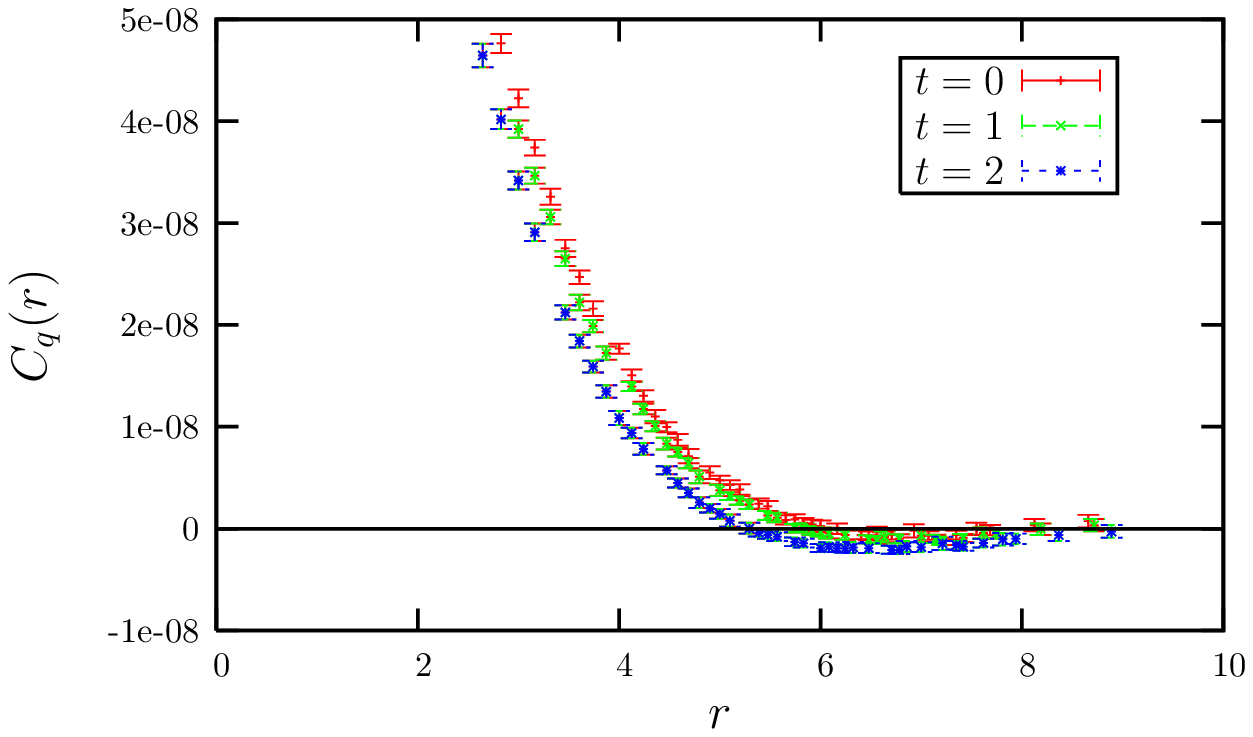,width=7.1cm}\\
$\kappa_{sea}=0.1343 \;(T<T_c)$ & $\kappa_{sea}=0.1360 \;(T>T_c)$
\end{tabular}
\end{center}
\caption{The topological charge correlator using 50 eigenmodes on 
${\Or}(200)$ configurations.}
\label{fig-correlator50}
\end{figure}

It is obvious that the UV cutoff in the topological density results
in a completely different structure actually seen by the lowest modes.
This is also reflected in Fig.~\ref{fig-correlator50} which shows the topological charge correlator Eq.~(\ref{eq-seiler}) in terms of the truncated density. We take 50 modes into account, which seems not to be sufficient to generate the negative tail of the correlator \cite{Koma:2005}. 
Since hypercubic rotation symmetry could be broken we plot the correlator
as a function of the Euclidean distance $r$  and the temporal distance $t$. 
The UV truncated correlator represents mainly size, shape and temperature
dependence of the clusters seen in the lower row of Fig.~\ref{fig-contour}. We see a slightly more negative correlation developing in the high-temperature phase.

\begin{figure}
\begin{center}
\hspace*{-0.5cm}\begin{tabular}{ccc}
\epsfig{file=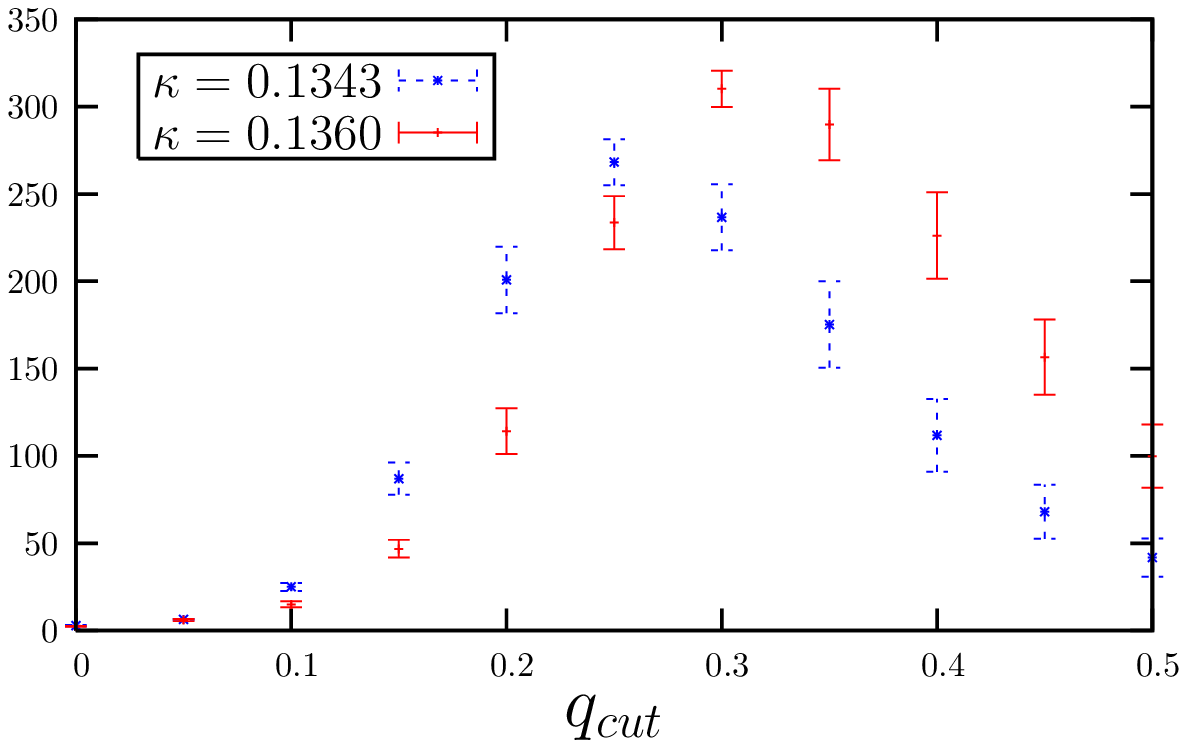, width=4.8cm}&
\epsfig{file=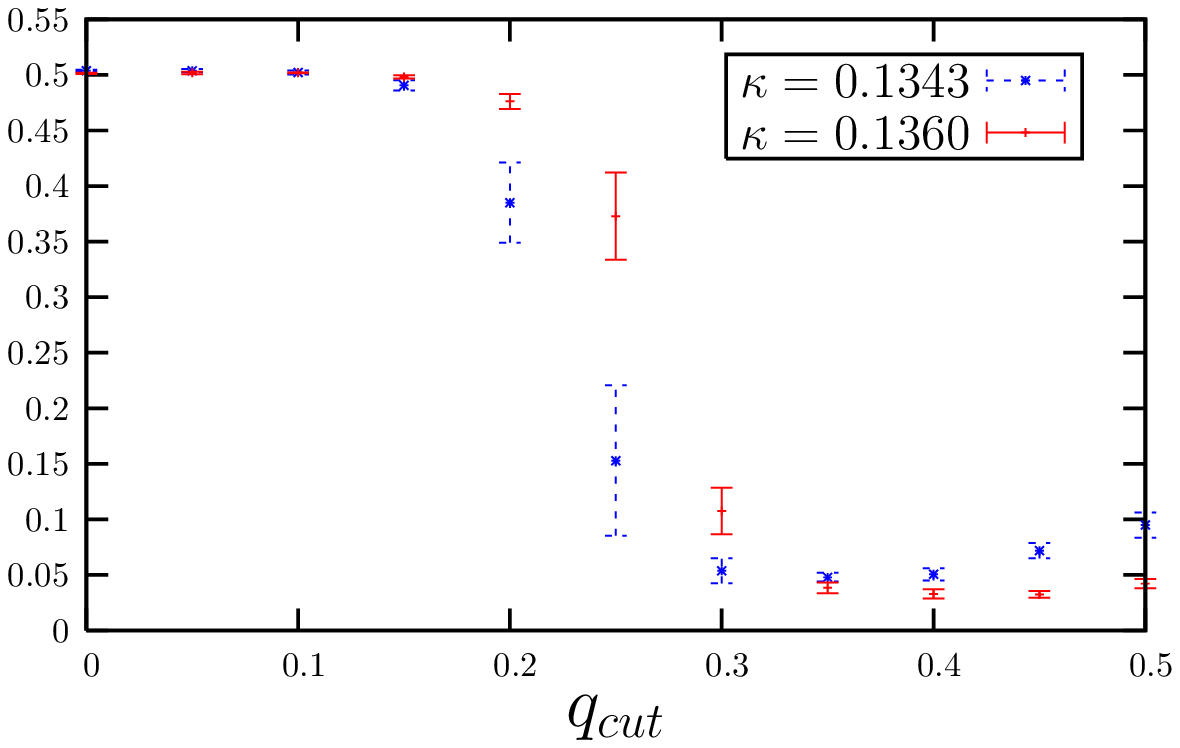, width=4.8cm}&
\epsfig{file=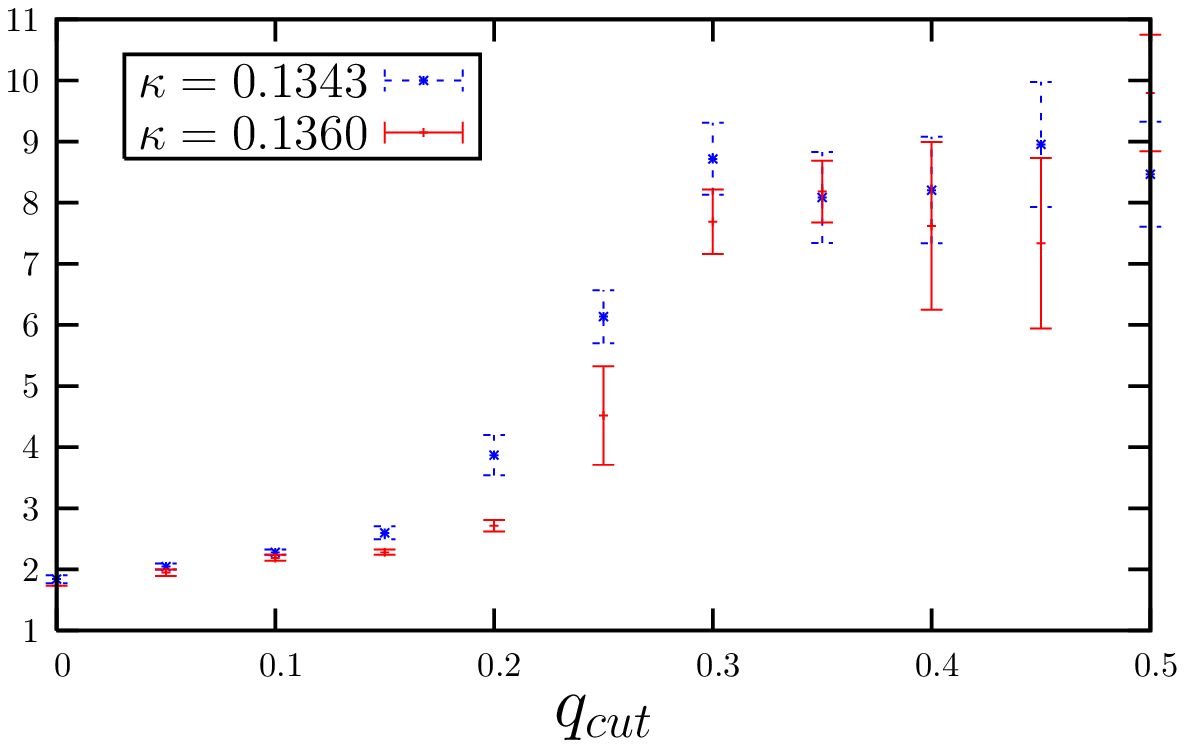, width=4.8cm}\\
(a) & (b) &  (c)\\
\end{tabular}
\caption{The total number of clusters of full density (a), the fractional size of the largest cluster (b) and the distance between the 2 leading clusters (c)   
depending on the cutoff value.}
\label{fig-cluster}
\end{center}
\end{figure}

\begin{figure}[h]
\begin{center}

\epsfig{file=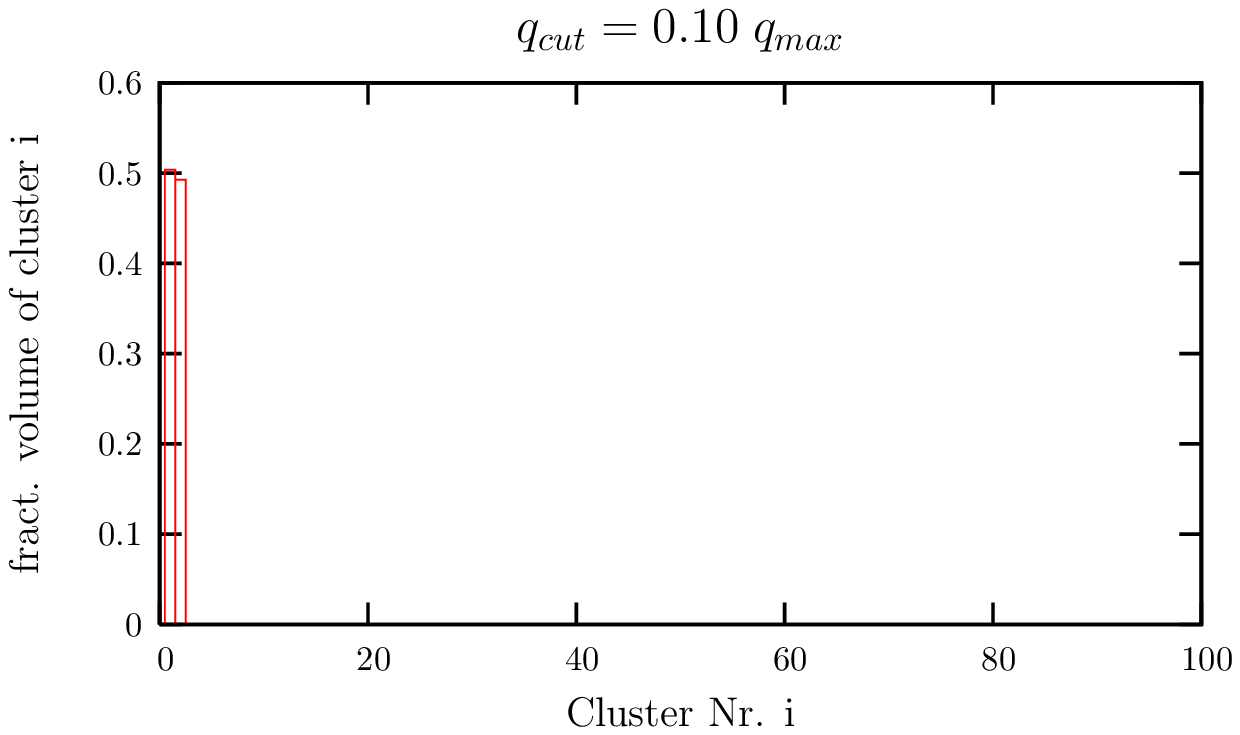,width=7cm}
\epsfig{file=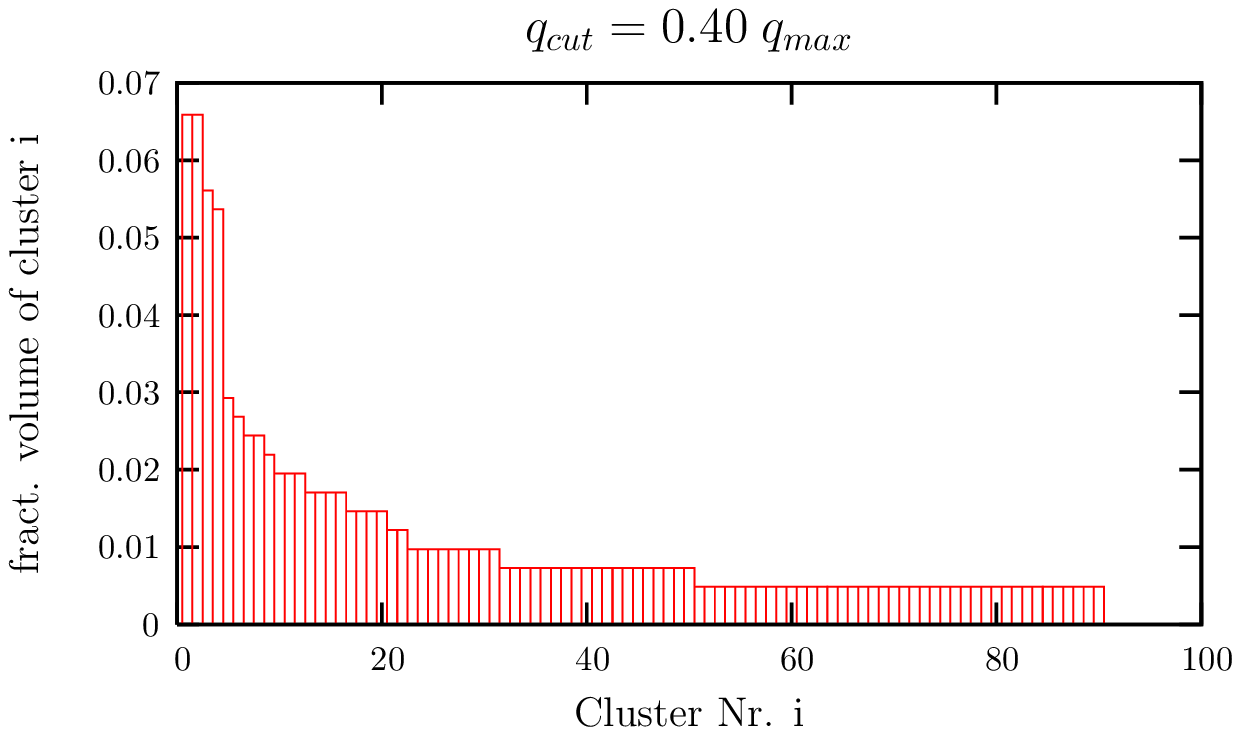,width=7cm}
\epsfig{file=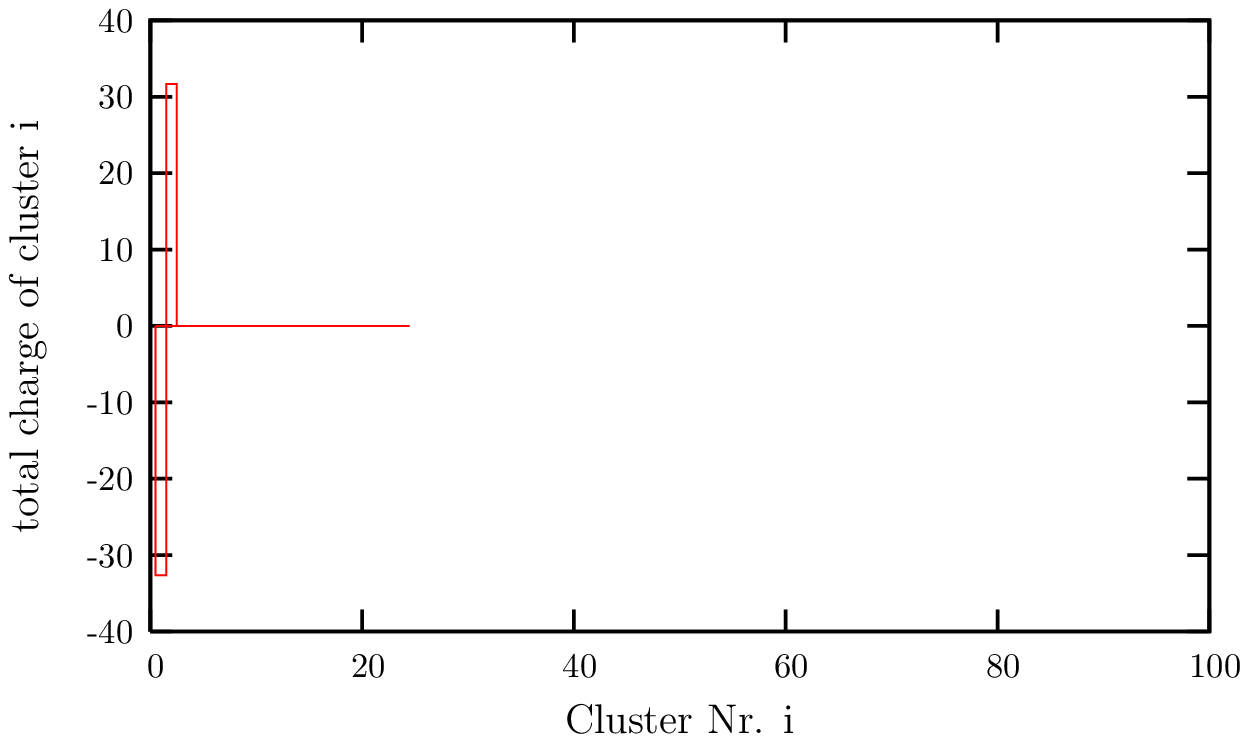,width=7cm}
\epsfig{file=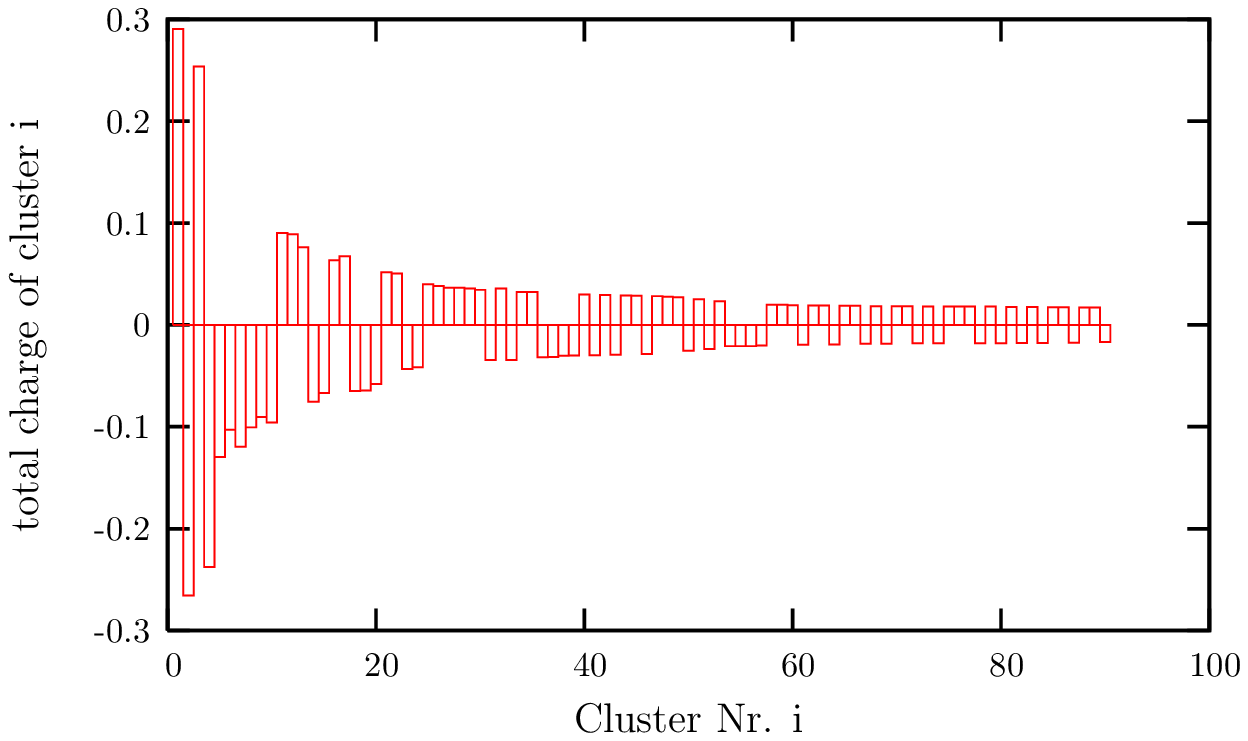,width=7cm}
\end{center}
\caption{Fractional volume (uppper plot) and total charge (lower plot) 
of the clusters of full charge 
sorted by their size for $q_{cut}=0.10\; q_{max}$ (left) and $0.40\; q_{max}$ 
(right) for one
configuration with $\kappa_{sea}=0.1343$.}
\label{fig-clustervol}

\end{figure}

Further investigating the full density, we show in  Fig.~\ref{fig-cluster} (a) and (b) that the number of clusters reaches a maximum at a moderate cutoff value, while for a 
cutoff as low as $q_{cut}=0.1\; q_{max}$
there are 2 totally dominating oppositely charged clusters with a fractional 
volume of around 50\% each.\footnote{The maximal number of clusters of the truncated density is an order of magnitude smaller.} 
The fractional volume and the total charge of 
the clusters ordered by their size are plotted in Fig.~\ref{fig-clustervol}.
Studying the 
point-to-point correlation of the largest clusters based on the full charge density, we find 
that the clusters start to percolate at $q_{cut}=0.2\; (0.25)\;q_{max}$ for $\kappa_{sea}=0.1343\; (0.1360)$.  Fig.~\ref{fig-cluster}~(c) shows the maximum of the minimal distance between each point of the largest  cluster and every point of the second largest one. 
The small value for low cutoffs in $q$ implies that the 2 dominating clusters 
are tangled and intertwined in a complex way.

A preliminary dimensional analysis indicates that we are rediscovering for the full density the picture of 2 dominating approximately 
3D sign coherent clusters \cite{Horvath:2005rv} 
at low cutoffs also at finite $T$, whereas we see 1D highly charged small clusters at high 
thresholds. 
This picture means that the QCD vacuum model with 4D coherent
(anti)instantons with a typical instanton radius of $0.3-0.4$ fm
is strongly modified by quantum fluctuations. The eigenmode-truncated density is more 
compatible with the conventional multiple-lump picture.
Apart from the topological susceptibilty which is strongly suppressed in the chiral-symmetry
restored phase, only quantitative structural differences between the phases
are seen so far.

\section*{Acknowledgements}

The numerical calculations have been performed at DESY-Zeuthen, 
LRZ Munich and the CIP Physik pool at the University of Munich. 
We thank these institutions for support. Part of this work is 
supported by DFG under contract FOR 465 
(Forschergruppe Gitter-Hadronen Ph\"anomenologie).

\bibliographystyle{JHEP-2}
\bibliography{proc}

\providecommand{\href}[2]{#2}\begingroup\raggedright\begin{thebibliography}{10}

\bibitem{Koma:2005}
Y.~Koma {\em et~al.}, {\it Localization properties of the topological charge
  density and the low lying eigenmodes of overlap fermions},  {\em
  PoS(LAT2005)300 (these proceedings)}
  [\href{http://arXiv.org/abs/hep-lat/0509164}{{\tt hep-lat/0509164}}].

\bibitem{Bornyakov:2004ii}
{\bf DIK} Collaboration, V.~G. Bornyakov {\em et~al.}, {\it Finite temperature
  QCD with two flavors of non- perturbatively improved wilson fermions},  {\em
  Phys. Rev.} {\bf D71} (2005) 114504
  [\href{http://arXiv.org/abs/hep-lat/0401014}{{\tt hep-lat/0401014}}].

\bibitem{Nakamura:2005}
Y.~Nakamura {\em et~al.}, {\it Critical temperature in QCD with two flavors of
  dynamical quarks},  {\em PoS(LAT2005)157 (these proceedings)}
  [\href{http://arXiv.org/abs/hep-lat/0509122}{{\tt hep-lat/0509122}}].

\bibitem{Galletly:2003vf}
{\bf QCDSF-UKQCD} Collaboration, D.~Galletly {\em et~al.}, {\it Quark spectra
  and light hadron phenomenology from overlap fermions with improved gauge
  field action},  {\em Nucl. Phys. Proc. Suppl.} {\bf 129} (2004) 453--455
  [\href{http://arXiv.org/abs/hep-lat/0310028}{{\tt hep-lat/0310028}}].

\bibitem{Aubin:2004mp}
{\bf MILC} Collaboration, C.~Aubin {\em et~al.}, {\it The scaling dimension of
  low lying Dirac eigenmodes and of the topological charge density},
  [\href{http://arXiv.org/abs/hep-lat/0410024}{{\tt hep-lat/0410024}}].

\bibitem{Hasenfratz:1998ri}
P.~Hasenfratz, V.~Laliena and F.~Niedermayer, {\it The index theorem in QCD
  with a finite cut-off},  {\em Phys. Lett.} {\bf B427} (1998) 125--131
  [\href{http://arXiv.org/abs/hep-lat/9801021}{{\tt hep-lat/9801021}}].

\bibitem{Horvath:2002yn}
I.~Horvath {\em et~al.}, {\it On the local structure of topological charge
  fluctuations in QCD},  {\em Phys. Rev.} {\bf D67} (2003) 011501
  [\href{http://arXiv.org/abs/hep-lat/0203027}{{\tt hep-lat/0203027}}].

\bibitem{Horvath:2003yj}
I.~Horvath {\em et~al.}, {\it Low-dimensional long-range topological charge
  structure in the QCD vacuum},  {\em Phys. Rev.} {\bf D68} (2003) 114505
  [\href{http://arXiv.org/abs/hep-lat/0302009}{{\tt hep-lat/0302009}}].

\bibitem{Seiler:2001je}
E.~Seiler, {\it Some more remarks on the Witten-Veneziano formula for the
  $\eta'$ mass},  {\em Phys. Lett.} {\bf B525} (2002) 355--359
  [\href{http://arXiv.org/abs/hep-th/0111125}{{\tt hep-th/0111125}}].

\bibitem{Horvath:2005cv}
I.~Horvath {\em et~al.}, {\it The negativity of the overlap-based topological
  charge density correlator in pure-glue QCD and the non-integrable nature of
  its contact part},  {\em Phys. Lett.} {\bf B617} (2005) 49--59
  [\href{http://arXiv.org/abs/hep-lat/0504005}{{\tt hep-lat/0504005}}].

\bibitem{Horvath:2005rv}
I.~Horvath {\em et~al.}, {\it Inherently global nature of topological charge
  fluctuations in QCD},  {\em Phys. Lett.} {\bf B612} (2005) 21--28
  [\href{http://arXiv.org/abs/hep-lat/0501025}{{\tt hep-lat/0501025}}].

\end{thebibliography}\endgroup

\end{document}